# Simultaneous X-Ray and optical spectroscopy of the Seyfert galaxy Mrk 993

A. Corral, X. Barcons, F.J. Carrera, M.T. Ceballos, and S. Mateos

Instituto de Física de Cantabria (CSIC-UC) 39005 Santander, Spain

**Abstract.** We present the results of simultaneous $XMM-Newton$ X-ray and 3.5m/CAHA optical spectroscopic observations of the Seyfert galaxy Mrk 993. This galaxy exhibited in the past significant variations of its broad emission lines which classified it as Sy1, Sy1.5 and Sy1.9 in different epochs. In order to test the X-ray/optical consistency of the unified AGN model we have compared the amount of photoelectric X-ray absorption with the Balmer decrement of broad emission line components. The optical data show that during our observations Mrk 993 was in a Sy1.8 state with a Balmer decrement of $\sim 9$. The X-ray absorbing gas amounts only to $7 \times 10^{20}$ cm$^{-2}$, most likely due to the host galaxy. For a normal gas to dust ratio in the absorbing material (expected to arise in the molecular torus) both quantities are clearly discrepant with an intrinsic Balmer decrement of $\sim 3.4$. We therefore conclude that the Balmer ratio is intrinsic rather than produced by obscuration and therefore that in this object the optical broad line emission properties are dictated by the physics of the broad line region rather than orientation/obscuration effects.

**Key words.** galaxies:active, galaxies:Seyfert, X-rays:galaxies

## 1. Introduction

In the framework of the unified model for Active Galactic Nuclei (AGN) there are two basic AGN types, AGN 1 and AGN 2 (Antonucci 1993). AGN 1 would present broad permitted and narrow forbidden emission lines in the optical and little (if any) absorption in the X-ray band as we are seeing the central engine face–on and, therefore, we see the Broad Line Region (BLR) unobscured by the molecular torus. On the contrary, AGN 2 should display significant absorption in the X-ray band, as we do not see the BLR but only the Narrow Line Region (NLR). This model predicts, of course, intermediate types depending on the inclination angle so that the strength of the broad permitted lines decreases as the absorption in the X-ray band increases. Such behaviour has been confirmed by many observations (see for example Smith & Done 1996, Nandra & Pounds 1994 and Bassani et al. 1999).

However, several recent observations have shown that this is not always true. On the one hand several authors (Mittaz et al. 1999, Fiore et al. 2000, Page et al. 2001, Schartel et al.2001, Tozzi et al. 2001, Mainieri et al. 2002, Brussa et al. 2003, Page et al. 2003, Perola et al. 2004 and Mateos et al. 2004) found AGN which present broad permitted lines (type 1 AGN) but are absorbed in X-rays. On the other hand, Pappa et al. (2001), Panessa et al. (2002), Barcons, Carrera & Ceballos (2003), Mateos et al. (2004) and Carrera, Page & Mittaz (2004), found AGN with weak or no broad emission lines and unabsorbed in X-rays, clearly in opposition to the simplest version of the AGN unified model, where broad line obscuration and X-ray absorption both arise in the same material – the torus.

Several explanations were suggested for the latest discrepancies (Pappa et al. 2001): the source might be Compton-thick so we only see scattered light; the central engine might be obscured in the optical by an ionised dusty absorber (with little effect on X–rays); and, finally, the difference is an intrinsic property, i.e., the properties of the broad line region are unrelated to orientation/absorption.

An additional possibility is that since the X-ray and optical observations are almost invariably non-simultaneous, they could map the same AGN in a different "absorption" state. That means that the amount of absorbing material (both in the X-rays and in the optical) would vary on scales from months to years. Variations in the optical spectral type on these scales have indeed been reported for a number of AGN (Aretxaga & Terlevich 1994). In that scenario, a simultaneous X-ray and optical observation would reveal an amount of X-ray photoelectric absorption consistent with the dust reddening of the BLR as mapped by the optical spectroscopy.

In order to test this last scenario, we performed a simultaneous X-ray and optical spectroscopic observation of the Seyfert galaxy Mrk 993 (MGC +05-04-058 / UGC 987, z = 0.015537, Huchra, Vogeley & Geller 1999). This galaxy has shown a high spectral type variability (Tran et al 1992) that moved it from type 1.5 to almost type 2 from its optical spectral characteristics. Further unpublished work by J.S. Miller and L.E. Kay (quoted by Tran et al.1992) revealed Seyfert 1 properties. In our observation, the source was caught in a Seyfert 1.8 optical state, but the amount of X-ray photoelectric absorption is clearly insufficient to explain this optical classification in terms of a standard (Balmer decrement ∼ 3.4) Seyfert 1 viewed through obscuring material.

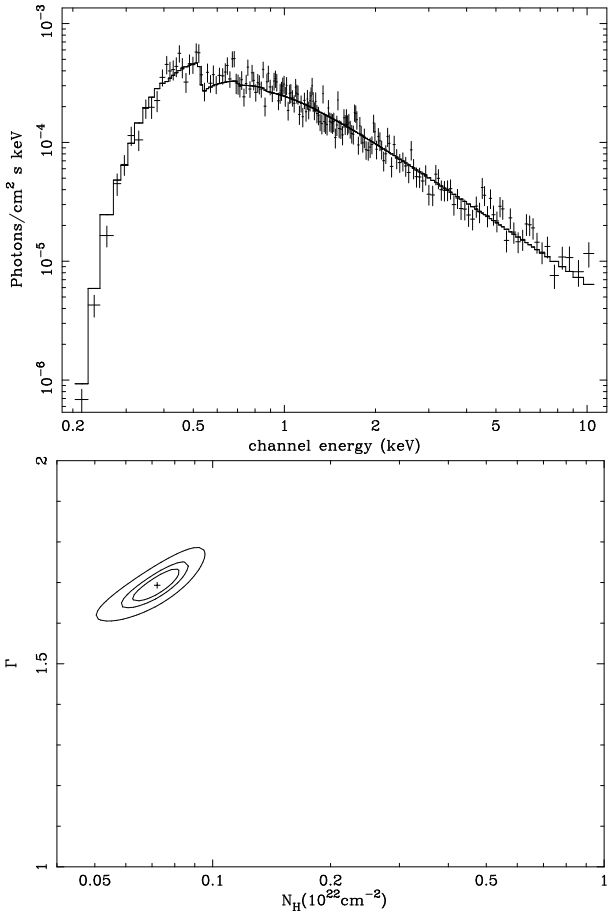

**Fig. 1.** EPIC–pn unfolded spectrum (top) and confidence contours $1\sigma$, $2\sigma$ and $3\sigma$ (bottom) for the best fitted model: pha ∗ zpha ∗ zpo. $N_H$ represents the intrinsic hydrogen column density and $\Gamma$ is the photon index.

This paper is organised as follows: in sections 2 and 3 we present the X-ray and optical data reduction and analysis respectively; in section 4 we compare the results in the two spectral bands; finally in section 5 we present our conclusions.

## 2. X-ray spectroscopy

Mrk 993 was observed with *XMM-Newton* during revolution 755 on the 23d of January of 2004 (Observation ID: 0201090401). The start and stop time of the observation were 16:53:53 and 23:12:26 UT. We used only the data from the EPIC instruments, MOS1, MOS2 and pn, which were operated in Full Window (imaging) mode with Thin1 filters. The exposure times for each instrument are 22.4, 22.4 and 20.8 ks for the MOS1, MOS2 and pn respectively. We did not use the RGS data because the spectra had very few source counts.

The data were pipeline-processed with the *XMM-Newton* Science Analysis Software (SAS) version 5.4.1. In all further work, including generation of calibration files, we used the most up-to-date version 6.0.0 of the SAS. The calibrated event lists were filtered in the time domain to avoid background flares. The resulting good time intervals were 15.8, 15.8 and 14.6 ks for the MOS1, MOS2 and pn respectively.

We extracted the source spectra in circular regions of 40 arcsec radii for all detectors. The background spectra were taken in circular source-free regions of 87 arcsec radii near the galaxy, avoiding CCD gaps. We selected single, double, triple and quadruple events in the MOS1 and MOS2 (pattern ≤ 12) and only single and double events in the pn (pattern ≤ 4). In the EPIC-pn spectra, only events with the highest spectral quality were included (i.e., `FLAG==0`). We generated redistribution and effective area matrices using SAS tasks `rmfgen` and `arfgen`. Finally, we binned the spectra so we have at least 30 counts per bin in order to use the $\chi^2$ statistic to fit the data.

The fitting was performed jointly to the MOS1, MOS2 and pn spectra with `XSPEC` (Arnaud et al. 1996) in the spectral range 0.2-12 keV. First, we fitted an absorbed power law with the hydrogen column density (model `phabs*zpowerlaw`) fixed at the Galactic value in the direction of Mrk 993 ($5.56 \times 10^{20}$ cm$^{-2}$, Dickey & Lockman 1990). That gave a poor fit ($\chi^2$/d.o.f. = 508/321). Then we added an intrinsic photoelectric absorption component (model `zphabs`), that resulted in a substantial improvement of > 99.99% in terms of the F-test ($\chi^2$/d.o.f. = 351/320).

The residuals of the fit corresponding to the pn only are suggestive of the presence of various emission lines (at ∼ 2, ∼ 4.6 and ∼ 6.4keV). We modelled each of these features using gaussian profiles but they are of < 80% significance in F-test terms, if we consider together the MOS1, MOS2 and pn, and at most ∼ 98%, if we only consider the pn. Given this very limited significance, we do not pursue the use of these features any further.

To search for an ionised absorber, we tried two models: `absori` (Done et al. 1992, Zdiarski et al. 1995), which provides a simple superposition of absorption edges dictated by the photoionisation state of the absorber, and the recent model `siabs` (Kinkhabwala et al. 2003) where both the absorption edge and the resonance absorption lines are included for a single ion species. In model `absori` we varied the absorber temperature, the overall chemical abundance and the ionisation parameter, and, in the model `siabs`, the absorber column density and the ion species (for C, S, Si, Mg and O). None of these models introduced significant improvements in the fit for any values of the parameter space explored. Specifically, for the model `absori`, the best fit gave a hydrogen column density of ∼ $7 \times 10^{20}$cm$^{-2}$ and an ionisation parameter of ∼ 0, i.e. we are led to the neutral absorber previously discussed. For the model `siabs` we looked for the spectral features that could arise in ionised species. Given the spectral range of our data the more representative absorption edges correspond to Oxygen, but the best fitted models (including the other ion species) gave a significance < 90% in the improvement of $\chi^2$ and correspond to nearly or totally neutral ions. For Oxygen ions we found that, a $3\sigma$ upper limit, for the ion column density ($N_{ion}$) is < $3 \times 10^{17}$ cm$^{-2}$. We therefore conclude that no significant ionised absorption is present in the source.

**Table 1.** Best fit spectral parameters corresponding to the model: `phabs*zphabs*zpowerlaw`

| Instrument | $\Gamma$ | $N_H$ ($10^{20}$ cm$^{-2}$) | Flux($10^{-12}$ erg cm$^{-2}$ s$^{-1}$)[a] |
|---|---|---|---|
| MOS | $1.69 \pm 0.07$ | $8.5 \pm 2.1$ | $1.35 \pm 0.02$ |
| pn | $1.71 \pm 0.06$ | $7.0 \pm 1.3$ | $1.33 \pm 0.02$ |
| MOS + pn | $1.69 \pm 0.04$ | $7.2 \pm 1.1$ | $1.34 \pm 0.02$ |

[a] Flux corresponding to the range 2-10 keV

The best fit model `phabs*zphabs*zpowerlaw` –a single power law at the source redshift photoelectrically absorbed by neutral gas at the same redshift as well as by a fixed amount of gas from our Galaxy– and the unfolded spectrum for the pn are represented in Fig. 1. The corresponding values to the fit are shown in Table 1.

## 3. The Optical Data

Mrk 993 was observed on 23d of January of 2004 with the 3.5 m telescope at the Calar Alto Observatory (CAHA). The first exposure was taken at 18:44:11 UT and the last one at 22:26:32 UT. We used the TWIN double spectrograph with central wavelengths of 4500 Å and 6700 Å for the blue and the red channels respectively. We obtained six exposures of 1000 s, three with the slit oriented in the direction of the major axis of the galaxy and the other three in the direction of the minor axis. We only considered in our analysis the three former exposures, as in this case we are able to subtract the light from the host galaxy (see later).

Using the standard IRAF long slit spectra reduction process and combining the three exposures, we obtained the spectra shown in Fig. 2. The reduction included debiasing, flat-fielding, wavelength calibration with arc lamps, and flux calibration using a standard star and the standard extinction curve for the observatory. The wavelength calibration (fitted to a fourth order polynomial) gave residuals of 0.02 Å for both the blue and the red spectra. The measured spectral resolution (FWHM of unblended arc lines) was 1.5 Å for the blue and 1.3 Å for the red. We propagated the errors along the whole process. Since the slit was not aligned with the parallactic angle ($\sim 30°$ apart), a certain amount of light may have been lost, preferentially in the blue. This probably affected the continuum shape so we have to be cautious about extracting directly information from it.

The spectral fitting was carried out using the QDP fitting routines via $\chi^2$ minimisation. A broad $H_\alpha$ component ($H_\alpha$(b)) is evident, in addition to narrow $H_\alpha$ ($H_\alpha$(n)) and N[II] doublet components. The $H_\beta$ broad component ($H_\beta$(b)), if any, is very weak. To search for it we fitted simultaneously the blue and the red spectra in the range 4800-5050 and 6500-6800 Å, tying down the velocity widths and redshifts of the broad $H_\alpha$ and $H_\beta$ to the same value. This wavelength range includes, in addition to the Balmer lines ($H_\alpha$ and $H_\beta$), the narrow emission lines [OIII] $\lambda 4958$ and the [NII] $\lambda\lambda 6548, 6583$ doublet. We fitted the spectrum with a linear component (to reproduce the continuum shape) plus a gaussian for each component line. The narrow and broad line components were fixed at the same redshift separately, as this improved the quality of the fit substantially. The [SII] $\lambda\lambda 6716, 6730$ doublet and the [OI] $\lambda 6300$ lines were fitted independently. The resulting Balmer decrements are $\sim 13$ for the NLR and $\sim 20$ for the BLR.

So far we have not considered the influence of the host galaxy in the resulting optical spectrum. Some authors (Mouri & Taniguchi 2002, Ivanov et al. 2000) have reported the possibility that Mrk 993 has a nuclear starburst. Ivanov et al. (2000) suggested, based on observations in the CO band, that Mrk 993 might have a circumnuclear starburst but they did not use it in their analysis because its spectrum had a low signal to noise ratio. Ivanov et al. (2000) classified Mrk 993 as 'host dominant', i.e., a galaxy that probably presents a circumnuclear starburst that hides the BLR. We also have to point out that the host galaxy is an emission line galaxy, displaying narrow emission lines such as $H_\alpha$ and the [NII] doublet, so the measured line intensities must be affected by this. Therefore, it is extremely important to obtain a host galaxy spectrum that represents the actual emission of the galaxy instead of using a template.

To take this into account we have extracted the spectrum of the host galaxy in order to subtract it from the obtained total (AGN+galaxy) spectrum as follows. First, for each exposure we extracted the AGN+galaxy spectrum and the galaxy spectrum from the same CCD image independently. We extracted the galaxy spectrum shifting the CCD image along the cross dispersion direction. Then we corrected the galaxy spectrum for the rotation curve (by shifting the wavelength direction to the nuclear frame) and combined all the galaxy spectra. We did not propagate the errors in this complicated process, but rather we estimated them by computing the standard deviation in a line-free region of the final spectrum. Once we had the AGN+galaxy and the galaxy spectra, we estimated the galaxy contribution using absorption lines from the host galaxy(e.g., G-band, Ca H) as a reference and subtracted the scaled resulting galaxy spectrum from the AGN+galaxy spectrum. The final spectra are shown in Fig. 3. Now the broad $H_\beta$ is much more evident. We performed the spectral fitting in the same way as we did with the non–subtracted spectra of Fig. 2, obtaining the line intensities listed on Table 2. In order to compute the errors in the Balmer decrement, we searched for the minimum $\chi^2$ for every fixed value of this parameter and all the others free. The measured values and the corresponding errors at 90% confidence level are the following:

$$\frac{H_\alpha}{H_\beta}(NLR) = 5.88 \pm 0.13, \qquad \frac{H_\alpha}{H_\beta}(BLR) = 8.99 \pm 0.02$$

The line at 5007Å of the [OIII] doublet was just missed by our setup. To compute it, we assumed a ratio of $\sim 3$ between this line and [OIII] $\lambda 4958$, a relation which holds with almost no exception (Osterbrock 1989). Using the Balmer decrement of the NLR to correct the narrow line intensities for reddening and following Baldwin, Phillips & Terlevich (1981), we found the following line ratios:

$$\log \frac{[OIII]\lambda 5007}{H_\beta} = 0.63, \quad \log \frac{[NII]\lambda 6583}{H_\alpha} = 0.51$$

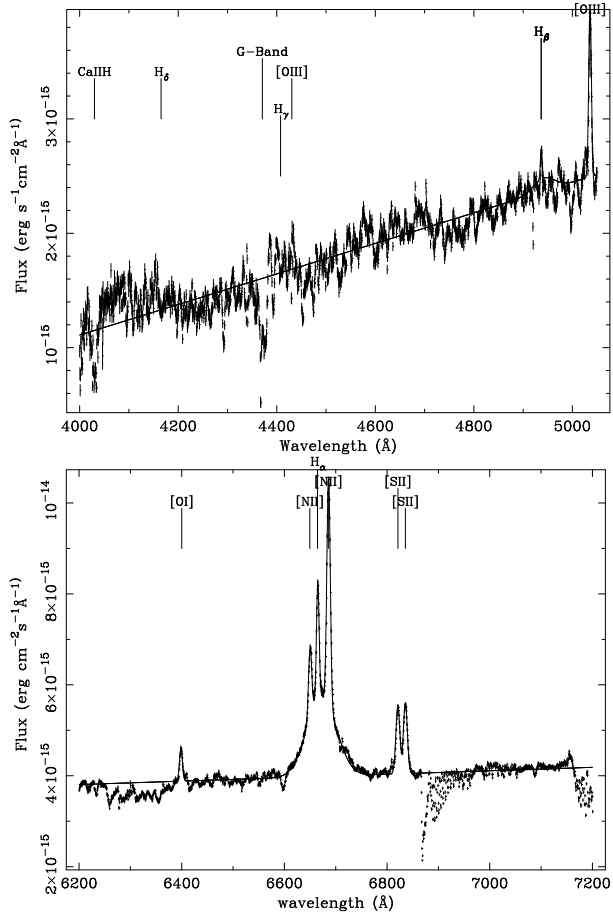

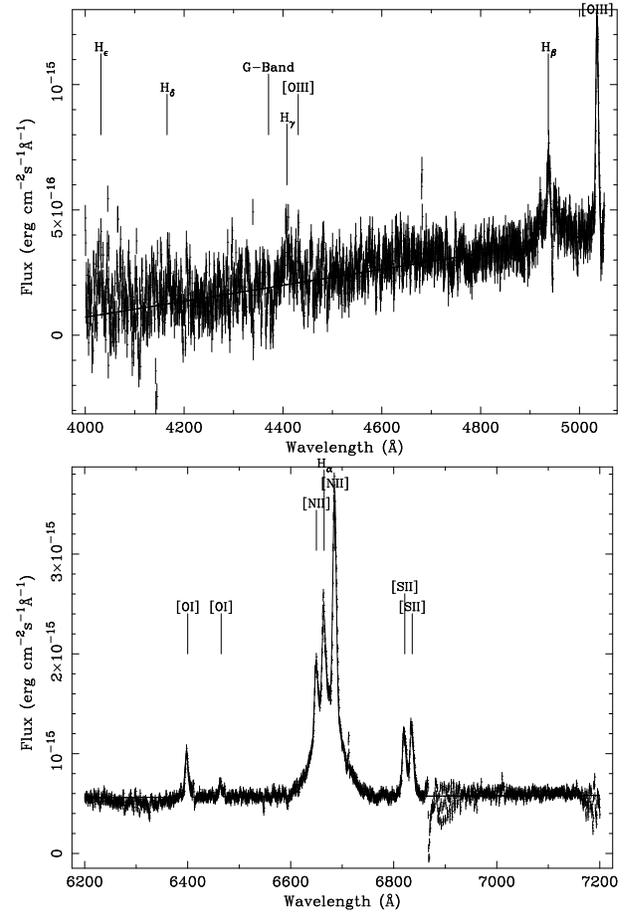

**Fig. 2.** Optical spectra error bars and model. Blue (top) and red (bottom)

**Fig. 3.** Galaxy subtracted optical spectra error bars and model. Blue (top) and red (bottom)

$$\log \frac{[S\,II]\lambda\lambda(6716+6730)}{H_\alpha} = 0.31\,, \quad \log \frac{[O\,I]\lambda 6300}{H_\alpha} = -0.19$$

These values place clearly the object in the AGN zone in line diagnostic diagrams (Osterbrock 1989).

**Table 2.** Line intensities and velocity widths with corresponding errors (galaxy subtracted spectrum). Errors are 90% confidence.

| Line | Intensity($10^{-15}$erg cm$^{-2}$ s$^{-1}$) | $\sigma$(km s$^{-1}$) |
|---|---|---|
| $H_\beta$(n) | 1.4 ± 0.2 | 164[a] |
| $H_\beta$(b) | 6.1 ± 1.1 | 1190[b] |
| [OIII]$\lambda$4958 | 6.1 ± 0.2 | 167 ± 8 |
| [OI]$\lambda$6300 | 3.6 ± 0.2 | 160 ± 10 |
| [OI]$\lambda$6365 | 1.0 ± 0.3 | 150 ± 50 |
| [NII]$\lambda$6548 | 6.06 ± 0.30 | 156 ± 6 |
| $H_\alpha$(n) | 8.40 ± 0.02 | 164 ± 5 |
| $H_\alpha$(b) | 60.5 ± 0.3 | 1190 ± 14 |
| [NII]$\lambda$6583 | 18.6 ± 0.2 | 155 ± 2 |
| [SII]$\lambda$6718 | 5.7 ± 0.3 | 160 ± 5 |
| [SII]$\lambda$6730 | 6.2 ± 0.2 | 160 ± 6 |

[a] Fixed at the same $\sigma$(km s$^{-1}$) as $H_\alpha$(n)
[b] Fixed at the same $\sigma$(km s$^{-1}$) as $H_\alpha$(b)

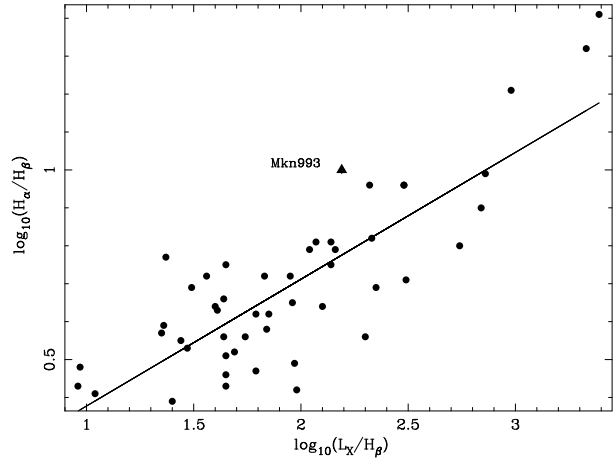

**Fig. 4.** Ward et al (1988) Fig. 4 and the fitted relation (solid line)

## 4. Discussion

We found that the Balmer decrement was ∼ 6 for the NLR. From this value we can estimate the hydrogen column density assuming a standard gas-to-dust-ratio (Bohlin, Savage & Drake 1978) obtaining $N_H \sim 3 \times 10^{21}$cm$^{-2}$, which, within errors, may be consistent with the summed absorption of the Galaxy and the host galaxy of Mrk 993. However, we found that the Balmer decrement of the BLR was ∼ 9 which, under the assumption of a *standard* case-B recombination and optically thin plasma Balmer decrement of ∼ 3.4 applicable to most Seyfert 1 galaxies, corresponds to a column density of $N_H \sim 5 \times 10^{21}$cm$^{-2}$. On the other hand, we derived a column density from the X-ray data of $N_H \sim 7 \times 10^{20}$cm$^{-2}$.

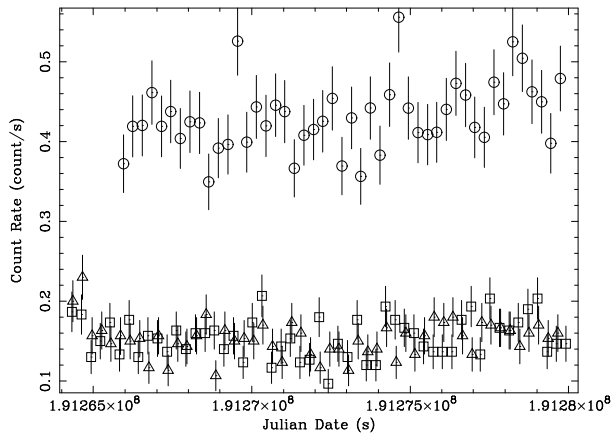

**Fig. 5.** Light curves for MOS1(squares), MOS2(triangles) and pn(circles) instruments, after background subtraction, corresponding to the 0.2-10 keV band.

Three possible explanations to this apparent discrepancy were suggested by Pappa et al. (2001): the galaxy is Compton-thick, there is a dusty warm absorber and finally, the obtained Balmer decrement is real and an intrinsic property of the BLR.

To examine the first possibility we tried to fit a Fe $K_\alpha$ line to the X-ray spectra. We only got a marginally significant (∼ 98%) detection of the line in the pn spectrum with an equivalent width of ∼270 eV. The measured flux in X-ray band 2-10 keV was $1.34 \times 10^{-12}$erg cm$^{-2}$s$^{-1}$. Even considering the "detection" of the Fe line as real and using the three dimensional diagnostic diagram proposed by Bassani et al. (1999), this small value versus the $F_X$ to F([OIII]$\lambda$5007) ratio (∼ 10) clearly rejects a Compton-thick scenario.

The second possibility was studied by searching for spectral features that must appear in the X–ray spectrum if a dusty warm absorber is present. As we saw in Sect. 2, we did not find evidence for an ionised absorber. Komossa & Bade (1998) predicted that a dusty warm absorber would smooth the oxygen edges but then we should see a C K-edge at 0.28 keV. There is nothing in our X-ray spectra that supports the presence of a significant absorption edge at 0.28 keV, although we cannot be completely sure since the EPIC–pn calibration it is not good enough below 0.3 keV.

The last possibility is that the Balmer decrement is an intrinsic property of the BLR. Ward et al. (1988) found a correlation between the Balmer decrement and the $F_X$ to F($H_\beta$) ratio for a large sample of Seyfert 1 galaxies. Despite the unlikely applicability of the case-B recombination and the likely optical thickness of the BLR clouds, they found an intrinsic Balmer decrement for the BLR of 3.5. They conclude that the differences in the Balmer decrements are due to nuclear reddening rather than being an intrinsic property of the BLR. However, when we represent Mrk 993 in that diagram (see Fig. 4) it has a Balmer decrement that place it formally $> 3\sigma$ above that correlation (solid line). If we compare this value with the Ward et al. (1988) sample, Mrk 993 has a deviation from the fit which is larger than at least 90% of the sample .

A similar result was found by Carrera, Page & Mittaz (2004) for the object RXJ133152.51+111643.5 and a much more extreme result was also found by Barcons, Carrera & Ceballos (2003) for the Seyfert 1.8/1.9 galaxy H1320+551. Both authors consider the possibility of non-simultaneity in the X-ray and optical observations as a possible way to resolve this discrepancy. This is certainly not our case, since our data were taken simultaneously so the results must represent the real absorption state of Mrk 993 at the time of the observation. We are therefore forced to conclude that the Balmer decrement of this source cannot be explained by obscuration but it has to be an intrinsic characteristic of the BLR.

There is, however, a caveat we have to consider. There could be a lag between changes in the ionisation spectrum – X-rays– and the optical emission lines due to the distance from the ionising source and structure of the BLR (recombination times might be significant). Our data extends only to hours while the expected lag could be of up to few days (see for example Dietrich et al. 1998), so our measurements of the X-ray variability cannot discard this possibility.

To estimate the source variability, we extracted the light curves for the pn and the MOS instruments. After correcting for the background we obtained the curves shown in Fig. 5. By fitting the data to a constant model, we found that variability is not significant (< 90%). Therefore, this result does not lend support the previous optical–lag explanation.

## 5. Conclusions

We have used a simultaneous X-ray and optical observation of the Seyfert galaxy Mrk 993 in order to test the X-ray absorption-optical obscuration relation predicted by the simplest version of the AGN unified model.

We found that, according to the optical data, the Mrk 993 nucleus is in a Sy 1.8 state with a Balmer decrement ∼9. However, the X-ray data implies little cold absorption giving an hydrogen column density of $< 10^{21}$cm$^{-2}$.

We have excluded the possibilities of the source being Compton-thick, and the presence of a dusty warm absorber to explain this difference. Since the observations are simultaneous, the spectral type variations observed for this source have been also excluded as explanation.

In short, unless the source varied significantly (for which we have no evidence) and there is a time lag between the variations of the ionising continuum and the state of the BLR, we have to conclude that Mrk 993 is not an obscured Sy 1 due to orientation effect but the Balmer decrement must be an actual characteristic of the BLR.

*Acknowledgements.* The work reported herein is based partly on observations obtained with XMM-Newton, an ESA science mission with instruments and contributions directly funded by ESA member states and the USA (NASA). It is also based on observations collected at the Centro Astronómico Hispano Alemán (CAHA) at Calar Alto, operated jointly by the Max-Planck Institut für Astronomie and the Instituto de Astrofísica de Andalucía (CSIC). Financial support for this work was provided by the Spanish Ministerio de Educación y Ciencia under project ESP2003-0812. AC acknowledges financial support from a Spanish Ministerio de Educación y Ciencia fellowship. We thank the referee, Jelle Kaastra, for useful suggestions.